\documentclass{article}
\usepackage{aaspp4,graphicx}
\begin{document}

\title{L1521E: A Starless Core in the Early Evolutionary Stage ?}
\author{Tomoya HIROTA}
\affil {Department of Physics, Faculty of Science, Kagoshima University,}
\affil {Korimoto 1-21-35, Kagoshima 890-0065, JAPAN ; 
hirota@sci.kagoshima-u.ac.jp}
\author{Tetsuya ITO, and Satoshi YAMAMOTO}
\affil {Department of Physics and Research Center for the Early Universe,}
\affil {The University of Tokyo, Bunkyo-ku, Tokyo 113-0033, JAPAN }

\begin{abstract}
We have studied the physical and chemical properties of 
a quiescent starless core L1521E with various molecular lines. 
It is found that 
there exists a compact dense core traced by the 
H$^{13}$CO$^{+}$, HN$^{13}$C, CCS, and HC$_{3}$N lines; 
their distributions have a single peak at the same position. 
The core radius is as small as 0.031 pc, 
whereas the H$_{2}$ density at the peak position is as high as 
(1.3-5.6)$\times$10$^{5}$ cm$^{-3}$. 
Although the density is high enough to excite the inversion 
transitions of NH$_{3}$, these lines are found to be very faint in 
L1521E. 
The distributions of NH$_{3}$ and CCS seem to be different 
from those of well-studied starless cores, L1498 and L1544, 
where the distribution of 
CCS shows a shell-like structure while that of NH$_{3}$ is 
concentrated at the center of the core. 
Abundances of carbon-chain molecules 
are higher in L1521E than the other dark cloud cores, 
and especially those of sulfur-bearing molecules C$_{n}$S 
are comparable to the cyanopolyyne peak of TMC-1. 
Our results suggest that 
L1521E would be in a very early stage of physical and chemical evolution. 
\end{abstract}


\section{Introduction}

On the basis of observations in various wavelength regions from radio to 
X-ray as well as theoretical considerations and simulations, 
it has been well established that dense cores in dark clouds are 
formation sites of low-mass stars. 
Because dense cores have low kinetic temperature (10 K) and 
high H$_{2}$ density ($>10^{3}$ cm$^{-3}$), 
they have been observed mainly with molecular lines. 
Especially, extensive survey observations of NH$_{3}$ 
by Myers and his collaborators have greatly contributed to understanding 
of the physical property of dark cloud cores 
(e.g. Benson \& Myers 1989). 
They found that 68 \% of the dense cores observed by the NH$_{3}$ 
lines ("NH$_{3}$ cores") accompany the {\it{IRAS}} sources, 
which are newly born stars in the cores (Benson \& Myers 1989; 
Beichman et al. 1986). Therefore, it has been recognized that the NH$_{3}$ 
lines are useful to study the physical properties of star-forming 
dense cores in dark clouds. 

On the other hand, 
Suzuki et al. (1992) pointed out that NH$_{3}$ is not always 
a good tracer of dense cores because of the chemical abundance 
variation from core to core. They carried out survey 
observations of CCS, HC$_{3}$N, HC$_{5}$N, and NH$_{3}$ 
toward 49 dense cores and found that the spectra of carbon-chain molecules 
tend to be intense in starless cores, 
while those of NH$_{3}$ tend to be intense in star forming cores. 
Especially, Suzuki et al. (1992) identified a few cores where the lines of 
carbon-chain molecules are intense while the NH$_{3}$ 
lines are hardly detected, which are called "carbon-chain-producing regions". 
They are L1495B, L1521B, L1521E, and 
the cyanopolyyne peak of TMC-1. 
Recently, Hirota, Ikeda, \& Yamamoto (2001) reported that 
the deuterium fractionation ratios of DNC/HNC and 
DCO$^{+}$/HCO$^{+}$ are significantly smaller in 
carbon-chain-producing regions than in the others. 
These systematic variation 
would reflect the difference in chemical evolutionary stages of the cores; 
carbon-chain molecules and NH$_{3}$ are abundant in relatively 
early and late stages, respectively (Suzuki et al. 1992) 
and deuterium fractionation ratios increase as the core evolves 
(Hirota et al. 2001; Saito et al. 2000). 

Although the dense cores identified by the NH$_{3}$ lines have extensively 
been observed with various molecular lines, no systematic study has been 
carried out for the carbon-chain-producing 
regions except for TMC-1 (e.g. Olano, Walmsley, \& Wilson 1988; 
Hirahara et al. 1992; Pratap et al. 1997). Detailed studies 
on the other carbon-chain-producing regions would be essential for understanding of 
chemical and physical evolution of dense cores. Among them, L1521E in the Taurus 
dark cloud complex (d=140 pc) is the most interesting source. It gives strong 
C$^{34}$S and CCS lines, whereas only weak NH$_{3}$ line has been detected so far. 
Particularly the C$^{34}$S emission is brightest among dense cores in our previous 
survey (Hirota et al. 1998). 
In this paper, we report the first comprehensive characterization of the chemical and 
physical properties of L1521E. 

\section{Observations}

The observed lines are summarized in Table \ref{tab-observe}. 
We took the reference position of L1521E to be 
($\alpha_{1950}, \delta_{1950}$)=
($04^{h}26^{m}12^{s}.5, 26^{\circ}$07\arcmin17\arcsec), 
where the spectra of 
H$^{13}$CO$^{+}$, HN$^{13}$C, CCS, and HC$_{3}$N
are found to be most intense. 
In the previous papers (e.g. Myers, Linke, \& Benson 1983), 
several molecular lines were observed 
at the position which is 30\arcsec \ north from our reference 
position. 

\placetable{tab-observe}

The 45 GHz and 76-93 GHz lines were observed with the 45 m radio 
telescope at Nobeyama Radio Observatory (NRO)\footnotemark 
\footnotetext{Nobeyama Radio Observatory is a branch of 
the National Astronomical Observatory of Japan, an interuniversity 
research institute operated by the Ministry of 
Education, Science, Sports and Culture of Japan} 
in several observing sessions from 1990 to 2000. 
All of them were observed with SIS mixer 
receivers whose system temperatures were about 200-500 K. 
The main-beam efficiencies ($\eta_{mb}$) were 0.7 and 0.5 for the 
45 GHz and 76-93 GHz regions, respectively, and the beam sizes 
were 37\arcsec \ and 20-17\arcsec \ for the 45 GHz and 76-93 GHz regions, 
respectively. 
Acousto-optical radio spectrometers 
with the frequency resolution of 37 kHz were 
used for the backend. 
Pointing was checked by observing a nearby SiO maser source, 
NML-Tau, every 1-2 hours, and the pointing accuracy was 
estimated to be better than 5\arcsec. 
All the observations were performed with the 
position-switching mode, in which a typical off position was 10\arcmin \ 
away from the source position. 

The 19-23 GHz lines 
were observed with the Effelsberg 100 m radio telescope of 
Max-Planck-Institut f\"ur Radioastronomie in 2000 May. 
We used cooled HEMT receivers, whose system temperatures 
were about 30 K. 
The digital autocorrelators with the frequency resolution of 10 kHz 
were used for the backend. 
For pointing and intensity calibrations, we observed 3C123 every 1-2 hours. 
We adopted a flux density of 3C123 to be 3.12 Jy 
at 23780 MHz (Ott et al. 1994). 
Observations were carried out with the frequency switching mode, 
in which the offset frequency was set to be 0.2 MHz. 

The $J=2-1$ lines of DNC and HN$^{13}$C were observed with 
the NRAO 12 m telescope\footnotemark 
\footnotetext{The National Radio Astronomy Observatory is a facility 
of the National Science Foundation, operated under cooperative agreement 
by Associated Universities, Inc.}
at Kitt Peak in 1999 December. We used two SIS mixer 
receivers and observed the DNC and HN$^{13}$C lines 
simultaneously. The system temperature was about 400 K. 
The main-beam efficiency is 
derived from the observations of planets to be 0.7 for both lines. 
The hybrid spectrometers with the frequency resolution of 48.8 kHz 
were used for the backend. 
We employed the position-switching mode as in the observations 
with the NRO 45 m telescope. 

\section{Dense Core in L1521E}

\begin{figure}
\begin{center}
\includegraphics[width=16cm]{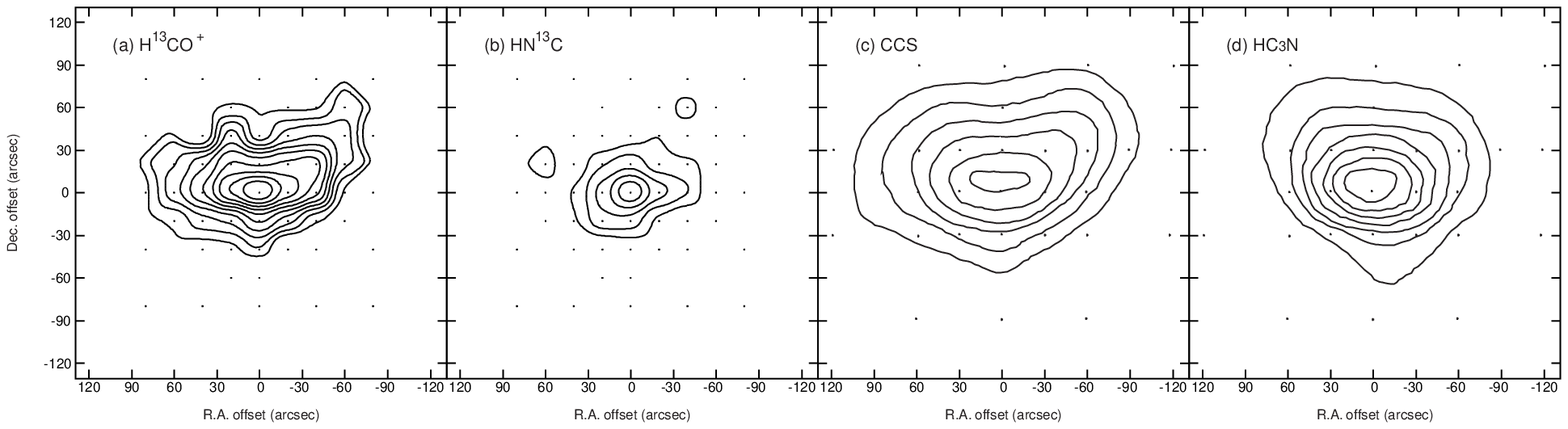}
\caption{Integrated intensity maps of the observed molecules. 
(a) H$^{13}$CO$^{+}$ ($J=1-0$). 
The velocity range of integration is from 4.5 to 7.0 km s$^{-1}$. 
The interval of the contours is 0.15 K km s$^{-1}$ and 
the lowest one is 0.3 K km s$^{-1}$. 
(b) HN$^{13}$C ($J=1-0$). 
The velocity range of integration is from 4.6 to 6.8 km s$^{-1}$. 
The interval of the contours is 0.08 K km s$^{-1}$ and 
the lowest one is 0.16 K km s$^{-1}$. 
(c) CCS ($J_{N}=4_{3}-3_{2}$). 
The velocity range of integration is from 6.5 to 6.9 km s$^{-1}$. 
The interval of the contours is 0.17 K km s$^{-1}$ and 
the lowest one is 0.17 K km s$^{-1}$. 
(d) HC$_{3}$N ($J=5-4$). 
The velocity range of integration is from 6.1 to 7.3 km s$^{-1}$. 
The interval of the contours is 0.29 K km s$^{-1}$ and 
the lowest one is 0.29 K km s$^{-1}$. 
\label{fig-maps}}
\end{center}
\end{figure}

Figure \ref{fig-maps} shows the integrated intensity maps of 
H$^{13}$CO$^{+}$, HN$^{13}$C, CCS, and HC$_{3}$N
in L1521E. 
It is clearly found that there exists a compact dense core traced by these 
lines, which has a single peak at the reference position with a structure elongated 
toward the east-west direction. 
The shape and peak position of these maps 
are similar to each other except for the slight difference in size, 
partly due to the difference in their critical density. There are neither signs 
of complex density variations nor of molecular abundance variations as seen
in TMC-1 (Hirahara et al. 1992, Pratap et al. 1997). 
The core size is 
derived to be 0.09 pc$\times$0.04 pc from the two dimensional Gaussian 
fitting of the integrated intensity map of H$^{13}$CO$^{+}$. 
The peak position of these lines is close to that 
of the C$^{18}$O line (Zhou et al. 1994), while the shape of the C$^{18}$O 
distribution is different from the above maps probably due to the lower critical density 
of C$^{18}$O. 

\placefigure{fig-maps}

\begin{figure}
\begin{center}
\includegraphics[width=12cm]{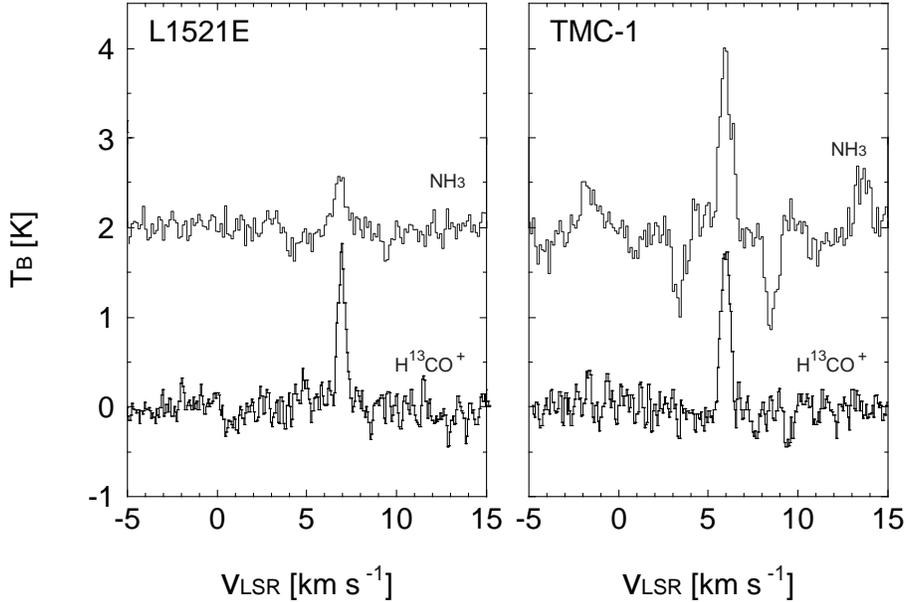}
\caption{Spectra of NH$_{3}$($J,K$=1,1) and 
H$^{13}$CO$^{+}$($J$=1-0) 
at the reference position of L1521E and the cyanopolyyne peak 
of TMC-1. The apparent features below the baseline of the NH$_{3}$ 
spectra are artifacts of the frequency switching technique. 
The hyperfine components of NH$_{3}$ can be seen at the 
$v_{lsr}$ of -12 km s$^{-1}$ and 13 km s$^{-1}$ for TMC-1. 
\label{fig-sp}}
\end{center}
\end{figure}

Figure \ref{fig-sp} shows the spectra of NH$_{3}$ ($J, K$=1,1) and 
H$^{13}$CO$^{+}$($J$=1-0) observed toward the reference position. 
The peak brightness temperature of the H$^{13}$CO$^{+}$ 
line is similar to those observed toward TMC-1 and other dense cores 
(Mizuno et al. 1994). In order to evaluate the 
H$_{2}$ density, we carried out the statistical equilibrium calculation based on 
the LVG model (Goldreich \& Kwan 1974) for DNC and HN$^{13}$C. 
The H$_{2}$ density derived from the $J$=1-0 and $J$=2-1 
lines of DNC is 4.5$\times10^{5}$ cm$^{-3}$, which is 
consistent with the density derived from the $J$=1-0 and 
$J$=2-1 lines of HN$^{13}$C (5.6$\times10^{5}$ cm$^{-3}$). 
These values are 
close to that derived from $J$=1-0 and $J$=2-1 lines of C$^{34}$S 
(1.3$\times10^{5}$ cm$^{-3}$; Hirota et al. 1998). 
The H$_{2}$ density of L1521E is comparable to those in other dense 
cores ( e.g. Benson \& Myers 1989; Hirota et al. 1998). 

\placefigure{fig-sp}

Although the H$_{2}$ density is high enough to excite the inversion 
line of NH$_{3}$($J, K$=1,1), it is found to be very faint toward 
the reference position of L1521E, as shown in Figure \ref{fig-sp}. 
Since the distribution of NH$_{3}$ is sometimes different from other molecules, 
we carried out the mapping observation of the NH$_{3}$ line toward the 
4\arcmin$\times$4\arcmin \ region centered at the reference position 
with spatial resolution of 40\arcsec. However, the NH$_{3}$ 
emission is generally weak over the entire core, and there is no other position where 
the NH$_{3}$ line is more intense than in the reference position. 
The distributions of CCS and NH$_{3}$ in L1521E seem to be different from those 
of the other starless cores previously observed, 
such as L1498 (Kuiper, Langer, \& Velusamy 1996) and L1544 
(Ohashi et al. 1999; Benson \& Myers 1989). In these cores, the distribution of 
CCS shows a shell-like structure, while that of NH$_{3}$ is concentrated 
at the central position. Such distributions can be interpreted qualitatively 
in terms of the chemical evolution; CCS tends to be depleted at the center of 
chemically evolved cores, whereas NH$_{3}$ becomes abundant there 
(Suzuki et al. 1992; Bergin \& Langer 1997). 
In contrast to L1498 and L1544, CCS is not depleted at the center of 
the core in L1521E, and NH$_{3}$ is less abundant there. Therefore, 
L1521E would be in the very early stage of its chemical evolution. 

It is notable that the core radius of L1521E, 0.031 pc, 
is significantly smaller than those of the other starless 
H$^{13}$CO$^{+}$ cores, 0.049-0.129 pc, and is rather comparable to 
those of the star-forming cores, 0.026-0.033 pc (Mizuno et al. 1994). 
Mizuno et al. (1994) interpreted this systematic difference 
in terms of the 
physical evolution of dense cores; star-forming cores are gravitationally more 
contracted than starless cores, and hence, the size of a dense part traced by the 
H$^{13}$CO$^{+}$ line would decrease from starless cores to star-forming 
cores. 
However, the L1521E core is apparently an exception to this scenario, 
because it is undoubtedly a starless core in spite of its small size; 
there are no {\it{IRAS}} point sources associated with L1521E. 
Furthermore, the submillimeter continuum emission and the evidence of 
molecular outflows have never been reported for L1521E. 
According to Lee, Myers, \& Tafalla (1999), the asymmetric line 
profile of CS($J$=2-1) indicating the infalling motion has not been detected 
toward the L1521E core. 
Considering the absence of intense NH$_{3}$ emissions, it is rather 
likely that L1521E is in a very early stage of physical evolution; 
the small size would indicate that L1521E 
is less evolved than the other starless cores, and hence, 
the high density region is still compact. 

The total mass of the H$^{13}$CO$^{+}$ core of L1521E is 
evaluated to be 2.4 $M_{\odot}$ assuming the peak density and the 
core radius of 3$\times$10$^{5}$ cm$^{-3}$ and 0.031 pc, respectively. 
The mass of the starless NH$_{3}$ core ranges from 
0.33 $M_{\odot}$ to 83 $M_{\odot}$ (Benson \& Myers 1989), 
and hence, the mass of L1521E corresponds to the lower end of the 
above range. This is consistent with the above result that 
L1521E is a still growing core. 
The virial mass of the H$^{13}$CO$^{+}$ core of 
L1521E is evaluated to be 
2.8 $M_{\odot}$ assuming the kinetic temperature of 10 K, 
which indicates that the core of L1521E would be gravitationally bound. 
In the maps of the CCS and H$^{13}$CO$^{+}$ lines, significant 
velocity gradient of 2 km s$^{-1}$ pc$^{-1}$ is observed
along the major axis; the velocity increases from east to west. 
If this gradient is ascribed to the rotation about the minor axis, 
the mass supported by centrifugal force is estimated to be only 
0.05 $M_{\odot}$. Therefore, the centrifugal force 
may contribute to cloud support only partly, as in the case of the 
NH$_{3}$ cores (Goodman et al. 1993). 

\section{Molecular Abundances}

\begin{figure}
\begin{center}
\includegraphics[width=8cm]{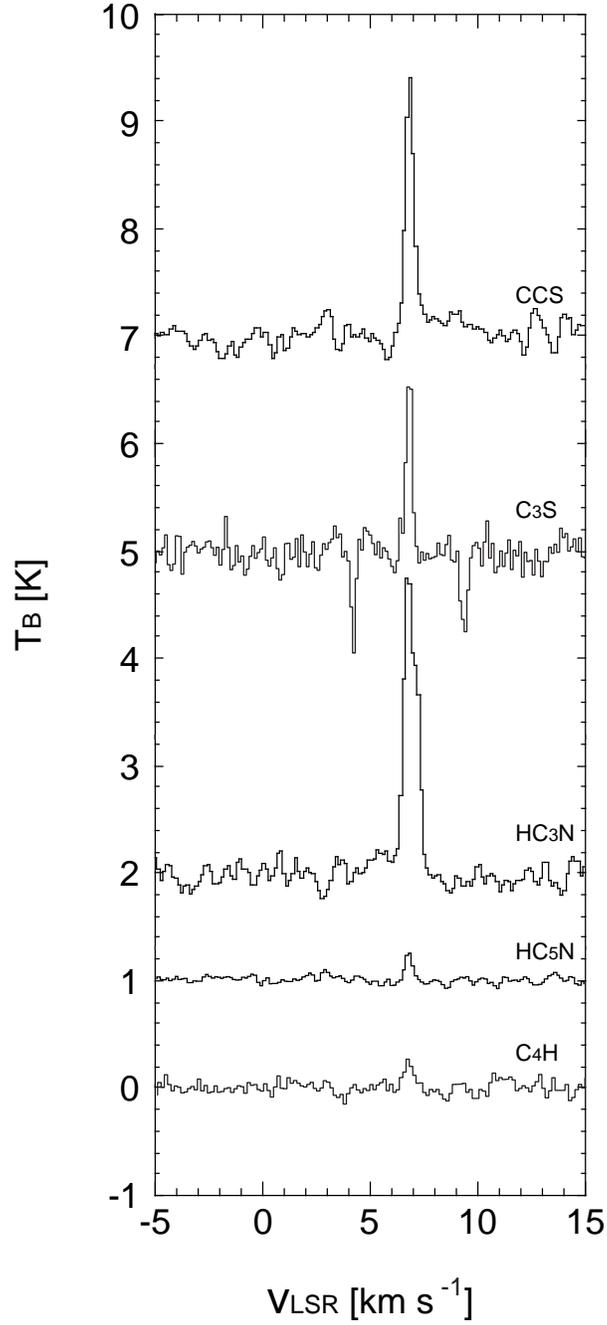}
\caption[]{Spectra of CCS, C$_{3}$S, HC$_{3}$N, 
HC$_{5}$N, and C$_{4}$H at the reference position of L1521E. 
The apparent features below the baseline of the C$_{3}$S 
spectrum are artifacts of the frequency switching technique. 
\label{fig-spcc}}
\end{center}
\end{figure}

We derived the abundances of various molecules toward the reference position. For 
this purpose, the line parameters for all the observed lines were determined by the 
Gaussian fit, as summarized in Table \ref{tab-observe}. 
We ignored the unresolved hyperfine 
structures of the NH$_{3}$, HC$_{3}$N, DNC, and HN$^{13}$C lines, 
and hence, their linewidths are broader than the others. 
In addition to the intense spectra of CCS and HC$_{3}$N, 
the longer carbon-chain molecules such as C$_{3}$S, C$_{4}$H, and 
HC$_{5}$N are also detected at the reference position as shown 
in Figure \ref{fig-spcc}. 
Therefore L1521E seems to be rich in carbon-chain molecules, as 
Suzuki et al. (1992) suggested. 

\placefigure{fig-spcc}

We calculated the column densities of observed molecules at the 
reference position of L1521E 
by the same method employed in 
Hirahara et al. (1992) and Suzuki et al. (1992) 
in order to compare the present results with those for 
the other dark cloud cores. 
We assumed the LTE condition with the excitation temperatures of 
6.5 K for NH$_{3}$, HC$_{3}$N, and HC$_{5}$N, 
6.0 K for C$_{4}$H, and 5.5 K for C$_{3}$S, 
because only a single line was observed each for these species. 
For NH$_{3}$, all the ortho and para levels 
were assumed to be thermalized at the kinetic temperature of 10 K. 
On the other hand, we carried 
out the LVG calculations for DNC and HN$^{13}$C. The column densities of DNC 
and HN$^{13}$C are derived by using the $J$=1-0 and $J$=2-1 transitions in 
the present study, and hence, they are reduced by a factor of 2-3 from those 
reported by Hirota et al. (2001), 
where the column density was determined only from the $J$=1-0 line with the 
assumption of the H$_{2}$ density of 1.3$\times10^{5}$ cm$^{-3}$. 
The derived column densities are 
summarized in Table \ref{tab-colmn} along with those of several 
molecules reported previously. 

It is important to estimate the optical depths of the observed lines, 
particularly for the NH$_{3}$ (1, 1) line. However, only the main 
hyperfine component was detected for the NH$_{3}$ (1, 1) line, 
so that we cannot determine the optical depth from the intensity ratio 
of the hyperfine components. Only the upper limit to the optical depth of 
the main hyperfine component can be estimated from the upper limit to the 
intensity of the strongest satellite component (i.e. three times the rms 
noise) to be 4.0. However the 
upper limit of the excitation temperature is evaluated as low as 3.3 K in this case. 
This excitation temperature is significantly lower than those reported for other dark 
cloud cores (4.5-7.5 K; Suzuki et al. 1992). Since the H$_{2}$ density of L1521E 
((1.3-5.6)$\times 10^{5}$ cm$^{-3}$) is 
enough high to excite the NH$_{3}$ (1, 1) line, the excitation temperature less 
than 3.3 K is unrealistic. When we assume the excitation temperature range 
from 4.5 K to 7.5 K observed toward other dark cloud cores, the column density 
determined ranges from 11.1$\times 10^{13}$ cm$^{-2}$ to 
6.7$\times 10^{13}$ cm$^{-2}$, whereas the optical depth ranges from 
0.39 to 0.13. Therefore, the reason why the NH$_{3}$ lines are hardly 
detected can be ascribed to the low column density. 

\placetable{tab-colmn}

In order to evaluate the fractional abundances of the observed molecules 
relative to H$_{2}$, 
we usually derive the column density of H$_{2}$ from that of C$^{18}$O 
because the fractional abundance of CO and its isotopic species relative to 
H$_{2}$ are well determined (Frerking, Langer, \& Wilson 1982). 
However, the distribution of C$^{18}$O in L1521E seems to be 
different from those of 
the other molecular lines as discussed above (Zhou et al. 1994) and hence, 
it is possible that the gas traced by the C$^{18}$O line is not the same as those 
traced by the observed lines of the other species 
whose critical densities are higher than 
that of C$^{18}$O. Pratap et al. (1997) suggested this point 
in their detailed study on TMC-1, and proposed to calculate the fractional 
abundances relative to HCO$^{+}$. 
Therefore, we also follow their analysis 
and compared the fractional abundances relative to H$^{13}$CO$^{+}$ 
between L1521E and TMC-1. 

The column densities of carbon-chain molecules toward L1521E 
are significantly higher than the typical NH$_{3}$ cores (Suzuki et al. 1992). 
Although the uncertainty of the calculated column densities are estimated 
to be a factor of a few mainly due to the assumed excitation temperature, 
the fractional abundances of C$^{34}$S, CCS, and C$_{3}$S relative to 
H$^{13}$CO$^{+}$ in L1521E are comparable to or higher than those of 
TMC-1, whereas those of longer carbon-chain molecules such as 
C$_{4}$H, HC$_{3}$N, and HC$_{5}$N are lower by an order of magnitude. 
On the other hand, the column density of NH$_{3}$ in L1521E is 
7.3$\times10^{13}$ cm$^{-2}$ which is apparently smaller 
by a factor of 10 
than those in the typical NH$_{3}$ cores
(Benson \& Myers 1989; Suzuki et al. 1992). 
In relation to this, we also found that the N$_{2}$H$^{+}$ line was not 
detected with the rms noise level of 0.2 K in L1521E, 
although it has been employed as a good tracer of dense cores (Lee et al. 1999). 
The upper limit of the column density of N$_{2}$H$^{+}$ is calculated 
by assuming the LTE condition with the excitation temperature of 5.0 K 
(Benson, Caselli, \& Myers 1998) and the FWHM linewidth of 0.5 km s$^{-1}$. 
We found that the fractional abundance of N$_{2}$H$^{+}$ relative to 
H$^{13}$CO$^{+}$ 
is at least smaller by a factor of 4 than that in the cyanopolyyne peak of 
TMC-1 (Hirahara et al. 1995), and hence, N$_{2}$H$^{+}$ would also be 
exceptionally deficient in L1521E. 

These characteristic features on molecular abundances seem to be understood 
as follows. Relatively high abundances of C$_{n}$S in comparison with 
TMC-1 may indicate that the gas-phase abundance of the sulfur atom might 
be higher in L1521E than in TMC-1 and other dense cores. 
One possible reason is that the heavy atoms are not so depleted in L1521E. 
This could be possible if the L1521E core is in 
the very early phase of its physical evolution (Bergin \& Langer 1997). 
In this case, the ionization degree would be enhanced, 
which gives an unfavorable condition for production of 
HC$_{3}$N and HC$_{5}$N, because the insertion reactions of the 
nitrogen atom to hydrocarbon ions compete with the electron 
recombination reactions. 
Furthermore the high degree of ionization provides an additional 
unfavorable situation for deuterium 
fractionation, which is consistent with the extremely low DNC/HNC 
and DCO$^{+}$/HCO$^{+}$ ratios observed in 
L1521E (Hirota et al. 2001). 
The low abundances of NH$_{3}$ and N$_{2}$H$^{+}$ directly 
indicate the young age of the L1521E core, since 
these species are only abundant in the late stage of chemical evolution. 

According to the physical properties and chemical abundances, 
L1521E could be a novel example of a very young core. Further 
detailed observations of this and similar cores would be of particular 
importance for understanding the initial phase of star formation processes. 

\acknowledgements

We are grateful to Masatoshi Ohishi, Shuji Saito, and Norio Kaifu 
for valuable discussions. 
We are also grateful to the staff of Nobeyama Radio Observatory, 
the NRAO 12 m telescope, and Effelsberg 100 m telescope of 
MPIfR for their assistance in observations. 
TH and TI thank to the Japan Society for the Promotion of Science 
for the financial support. 
This study is partly supported by Grant-in-Aid from Ministry of 
Education, Science, Sports and Culture of Japan 
(07CE2002 and 11304010).

{}
\newpage

\newpage


\begin{deluxetable}{lrccccccc}
\tablenum{1}
\tablewidth{0pt}
\tablecaption{Observed lines 
\label{tab-observe}}
\tablehead{
\colhead{} & 
  \colhead{$\nu$} &   \colhead{} & \colhead{$\mu$\tablenotemark{b}} & 
  \colhead{} & \colhead{$T_{B}$} &   $v_{lsr}$ & 
  \colhead{$\Delta v$} &  \colhead{$T_{rms}$}  \nl
\colhead{Transition} & 
  \colhead{(MHz)} & \colhead{$S_{ul}$\tablenotemark{a}} & 
  \colhead{(Debye)} & 
  \colhead{Telescope} & \colhead{(K)} & \colhead{(km s$^{-1}$)} &
  \colhead{(km s$^{-1}$)} &  \colhead{(K)} 
}
\startdata
C$_{4}$H($N$=2-1,$J$=$\frac{5}{2}$-$\frac{3}{2}$,$F$=3-2)  
 & 19015.144 & 2.00 & 0.9 & 
  MPIfR &  0.24  & 6.78 & 0.48 & 0.06 \nl
NH$_{3}$($J, K$ = 1,1)  & 23694.506\tablenotemark{c}  & 1.50 & 1.47 & 
  MPIfR & 0.57  & 6.84 &  0.72 &  0.13 \nl
C$_{3}$S($J$=4-3)  & 23122.985  & 4.00 & 3.6 & 
  MPIfR & 1.71  & 6.81 &  0.32 & 0.17 \nl
HC$_{5}$N($J$=16-15)  & 42602.153  & 16.00 & 4.33 & 
  NRO & 0.39  & 6.79 &  0.39 & 0.06 \nl
CCS($J_{N}$=$4_{3}$-$3_{2}$)   & 45379.033  & 3.97 & 2.81 & 
  NRO & 2.5  & 6.83 &  0.42 &  0.1  \nl
HC$_{3}$N($J$=5-4)  & 45490.316\tablenotemark{c}  & 5.00 & 3.72 & 
  NRO & 2.7  & 6.86 &  0.73 & 0.1 \nl
DNC($J$=1-0)  & 76305.727   &  1.00  & 3.05 & 
  NRO &  0.34  & 6.87 &  0.78 & 0.09  \nl
DNC($J$=2-1)   & 152609.774   &  2.00  & 3.05 & 
  NRAO & 0.18  & 6.87 &  0.54 & 0.04 \nl
HN$^{13}$C($J$=1-0)  & 87090.850  & 1.00  & 3.05 & 
  NRO &   0.69  & 6.91 &  0.62 & 0.09  \nl
HN$^{13}$C($J$=2-1)  & 174179.411   &  2.00  & 3.05 & 
  NRAO & 0.31  & 6.77 &  0.42 & 0.06  \nl
H$^{13}$CO$^{+}$($J$=1-0)  & 86754.330  & 1.00 & 4.07 & 
  NRO & 1.72  & 6.93 &  0.52 &  0.14  \nl
N$_{2}$H$^{+}$($J$=1-0) & 93173.777\tablenotemark{c} & 3.40 & 
   1.00 & NRO & \nodata & \nodata &  \nodata & 0.2 \nl
\enddata
\tablenotetext{ a}{ Intrinsic line strength}
\tablenotetext{ b}{ Dipole moment}
\tablenotetext{ c}{ Main hyperfine component}
\end{deluxetable}

\begin{deluxetable}{lcccc}
\tablenum{2}
\tablewidth{0pt}
\tablecaption{Column densities of the selected molecules 
in unit of $10^{13}$ cm$^{-2}$
\label{tab-colmn}}
\tablehead{
\colhead{Molecule} & 
  \colhead{L1521E} &   \colhead{Reference} & 
  \colhead{TMC-1\tablenotemark{a}} &  \colhead{Reference} 
}
\startdata
C$^{34}$S          &     1.25    & 1 &   0.73      &  1  \nl
CCS                     &     2.8   & 2  &  6.6      & 3 \nl
C$_{3}$S           &     1.4    & \nodata & 1.3      & 3 \nl
C$_{4}$H           & 9.8   & \nodata & 440 & 4 \nl
HC$_{3}$N        &     2.3    & \nodata  &  17.1     & 3 \nl
HC$_{5}$N       &     0.46  & \nodata &  5.6      &  3  \nl
NH$_{3}$          &     7.3   & \nodata & 19       & 3 \nl
N$_{2}$H$^{+}$ & $<$0.14   & \nodata & 0.74     & 5 \nl
DNC                   &     0.042  & \nodata &  0.89     & 2 \nl
HN$^{13}$C     &     0.064  & \nodata &  0.53     & 2 \nl
H$^{13}$CN     &      0.23  & 1 &   0.37      &  1 \nl
H$^{13}$CO$^{+}$ & 0.106   & 2 & 0.14     & 6 \nl
C$^{18}$O       &      170    & 7 &  330        &  6 \nl
\enddata
\tablenotetext{a}{The position is cyanopolyyne peak except for 
C$_{4}$H, which is at 1\arcmin southeast of the cyanopolyyne peak. }
\tablerefs{1: Hirota et al. (1998); 2: Hirota et al. (2001); 
  3: Suzuki et al. (1992); 4: Hirahara et al. (1992); 
  5: Hirahara et al. (1995); 
  6: Pratap et al. (1997); 7: Myers et al. (1983)}
\end{deluxetable}


\end{document}